\documentclass[12pt,usenames,dvipsnames]{article}

\usepackage{latexsym}
\usepackage{amssymb,amsfonts,amsmath}
\usepackage{graphicx} 
\usepackage{indentfirst}
\usepackage{bbm}
\usepackage{amssymb}
\usepackage{verbatim}
\usepackage{amsmath, amsthm,amssymb}
\usepackage{mathrsfs}
\usepackage{hyperref}
\usepackage{amsfonts}
\usepackage{dsfont}
\usepackage{cite}
\usepackage{xcolor}
\usepackage{enumerate}
\usepackage{cleveref}
\usepackage{braket}

\usepackage{tensor}

\parskip=\medskipamount

\newcommand {\cD}{{\cal D}}

\newcommand {\cK}{{\cal K}}
\newcommand {\cL}{{\cal L}}

\newcommand {\cN}{{\cal N}}

\def\a{\alpha}

\def\b{\beta}
\def\c{\chi}
\def\d{\delta}
\def\e{\epsilon}

\def\g{\gamma}

\def\l{\lambda}

\def\o{\omega}

\def\s{\sigma}

\def\x{\xi}

\def\D{\Delta}

\def\J{\Psi}
\def\L{\Lambda}

\def\tr{{\rm tr}}
\def\rd{{\rm d}}
\def\ri{{\rm i}}
\def\re{{\rm e}}

\newcommand{\ve}{\varepsilon}                            %

\newcommand{\pa}{\partial}                           %
\newcommand{\hf}{\frac12}

\newcommand{\bea}{\begin{eqnarray}}
\newcommand{\eea}{\end{eqnarray}}
\newcommand{\non}{\nonumber}

\newcommand{\bm}[1]{\mbox{\boldmath$#1$}}

\def\double #1{#1{\hbox{\kern-2pt $#1$}}}

\newif\ifdtup

\newcommand{\bsubeq}{\begin{subequations}}
\newcommand{\esubeq}{\end{subequations}}

\numberwithin{equation}{section}

\newcommand{\sSO}{\mathsf{SO}}

\newcommand{\sU}{\mathsf{U}}

\begin{document}

\begin{titlepage}
\begin{flushright}
April 2026\\
Revised version: November 2026
\end{flushright}
\vspace{5mm}

\begin{center}
{\Large \bf 
Euler-Heisenberg actions in higher dimensions
}
\end{center}

\begin{center}

{\bf Terry Hatzis and Sergei M. Kuzenko} \\
\vspace{5mm}

\footnotesize{ 
{\it Department of Physics M013, The University of Western Australia\\
35 Stirling Highway, Perth W.A. 6009, Australia}}  
~\\
\vspace{2mm}
~\\
Email: \texttt{terry.hatzis@uwa.edu.au,
sergei.kuzenko@uwa.edu.au}\\
\vspace{2mm}

\end{center}

\begin{abstract}
\baselineskip=14pt
 
Using Schwinger's proper-time formalism, we compute the one-loop effective action in both spinor and scalar quantum electrodynamics in $d=2n>4$  dimensions using algebraic techniques. We derive the closed form expression for the corresponding Euler-Heisenberg Lagrangian, and extract its weak-field approximation in 6, 8 and 10 dimensions in terms of invariants of the  electromagnetic field. A subsequent analysis of pair production in $d$ dimensions is given. We also elaborate on conformal actions for the electromagnetic field in $d=2n>4$ dimensions. In the $d=6$ and 8 cases, such actions are generated by DeWitt's coefficients $a_3$ and $a_4$ which, for completeness, are re-computed in this work for spinor QED and scalar QED, respectively. 
\end{abstract}
\vspace{5mm}

\vfill

\vfill
\end{titlepage}

\newpage
\renewcommand{\thefootnote}{\arabic{footnote}}
\setcounter{footnote}{0}

\tableofcontents{}
\vspace{1cm}
\bigskip\hrule

\allowdisplaybreaks

\section{Introduction}

Ten years ago, the low-energy effective (LEE) action for the $q$-hypermultiplet 
coupled to a background vector multiplet in six dimensions (6D) was computed in ${\cal N}=(1,0)$ harmonic superspace
\cite{Buchbinder:2016acg}, building on the results of earlier publications 
\cite{Buchbinder:2015tea, Buchbinder:2014sna}. One may think of the construction given in \cite{Buchbinder:2016acg}
as a massless 6D supersymmetric extension of the Euler-Heisenberg action
\cite{Heisenberg:1936nmg, Weisskopf:1936hya}, see \cite{Dunne:2004nc, Dunne:2012vv} for a review.\footnote{In the case of 6D ${\cal N}=(1,0)$ supersymmetry, the hypermultplet is massless.} 
The superfield LEE  action derived in \cite{Buchbinder:2016acg} has not been reduced to components, while the explicit component form of the LEE action is certainly of interest. It is quite surprising that, to the best of our knowledge, 
a six-dimensional analogue of the Euler-Heisenberg effective action has not appeared in the literature, for both spinor and scalar quantum electrodynamics (QED) in six dimensions. One of the aims of this paper is to fill the gap.

Modern approaches to computing various LEE actions, including the Euler-Heisenberg action, go back to the 1951 work by Schwinger \cite{Schwinger:1951nm} and its generalisation to curved space by DeWitt \cite{DeWitt:1964mxt, DeWitt:2003pm}. For a recent review of the Schwinger-DeWitt technique, see, e.g., \cite{Barvinskii:2024iqz}.
Superfield variants of the Schwinger-DeWitt technique have been used to compute $\cN=1$ and $\cN=2$ supersymmetric 
analogues of the Euler-Heisenberg action in four dimensions
\cite{Shizuya:1983pn, McArthur:1997ww, Pletnev:1998yu, Buchbinder:1999jn, Kuzenko:2003qg}.
These results have been extended to two loops first for $\cN=2$ supersymmetric QED (SQED) \cite{Kuzenko:2003qg}, and then to $\cN=1$ SQED \cite{Kuzenko:2007cg} in four dimensions, see \cite{Tyler:2013mgu} for a review.\footnote{Two-loop LEE actons on $\cN=2$ and $\cN=4$ three-dimensional SQED were computed in \cite{Buchbinder:2013jca}.}

In six dimensions, it is well known that the standard spinor QED with Lagrangian
\bea
L = - \frac{f^2}{4} F^{ab}F_{ab} + \bar{\Psi}(\ri \gamma^a \nabla_a-m)\Psi
\eea
is not renormalisable, where $f$ is the coupling constant of mass dimension $+1$.  However, one can consider its higher-derivative extension
\bea
\widetilde{L} =  \frac{1}{2g^2} \pa_bF^{ab} \pa^c F_{ac} 
- \frac{f^2}{4} F^{ab}F_{ab} + \bar{\Psi}(\ri \gamma^a \nabla_a-m)\Psi~, 
\label{renormalisable}
\eea
with $g$ the dimensionless coupling constant. This theory is renormalisable, and its version with $f=m=0$ is classically conformal. Renormalisable higher-derivative supersymmetric Yang-Mills theories in six dimensions were introduced by 
Ivanov, Smilga and Zupnik \cite{Ivanov:2005qf}, 
and their ultraviolet behaviour was thoroughly studied in \cite{Bossard:2015dva}.\footnote{The six-dimensional $\cN= (1,0)$ supersymmetric $F \Box F$ theory \cite{Ivanov:2005qf}  was coupled to $\cN=(1,0)$ conformal supergravity in \cite{Butter:2016qkx}.}
Conformal Yang-Mills theories in $d=2n>6$ dimensions were constructed by Metsaev in Minkowski \cite{Metsaev:2023qif}
and anti-de Sitter \cite{Metsaev:2023qif} space, and their universal feature is the presence of higher derivatives. Using such actions and their lower-derivative cousins allows one to introduce renormalisable analogues of \eqref{renormalisable} in $d=2n>6$ dimensions. Beyond six dimensions, $n>3$, the Lagrangian should also include self-interaction terms ${\rm tr} \big(F^{2k}\big)$, with $1<k \leq \lfloor n/2 \rfloor$. 

This paper is organised as follows. Section \ref{section2}  contains a brief review of Schwinger's proper-time formalism and its extension to $d$ dimensions, and the six dimensional Euler-Heisenberg action is derived in closed form. In Section \ref{section3} we present the closed form expression of the Euler-Heisenberg action in $d=2n$ dimensions. In Section \ref{section4}, we use the obtained results to derive the pair production rate in $d=2n$ dimensions. 
In Section \ref{section5}, on conformal actions for the electromagnetic field in $d=2n>4$ dimensions.
Discussion and conclusions are given in Section \ref{section6}. The main body of the paper is also supplemented by three appendices. Appendix \ref{appendixA} reviews the relevant aspects of the spinor formalism in $d=2n$ dimensions.  
Appendix \ref{appendixB} discusses the structure of the invariants of the electromagnetic field in six dimensions. Appendix \ref{appendixC} is devoted to the calculation of DeWitt's coefficients $[a_3]$ and $[a_4]$ in flat spacetime.


\section{One-loop effective action in six dimensions}\label{section2}

In this section we use the proper-time approach \cite{Schwinger:1951nm,DeWitt:1964mxt}
(see also \cite{Dittrich:1985yb,Dittrich:2000zu, Avr, Vassilevich:2003xt}) to compute the one-loop effective action $\Gamma^{(1)}[A]$ of spinor QED
in six dimensions in closed form. 
In $d$ dimensions, the Dirac spinor field coupled to an electromagnetic field is described by the action
\bea
    S[\Psi,\bar{\Psi};A] = \int \rd^d x\,\bar{\Psi}(\ri \gamma^a \nabla_a-m)\Psi\,,
\eea
where $\nabla_a = \partial_a -\ri eA_a$ is the $\sU{(1)}$ gauge-covariant derivative. 

We also compute the analogous result for the case of a charged scalar field $\phi$ coupled to a background electromagnetic field, with action
\bea
    S[\phi,\phi^*;A] = \int \rd^d x\, \left[-(\nabla^a \phi)(\nabla_a\phi)^*-m^2\phi\phi^* \right].
\eea
For both effective actions, we compute the relevant weak-field expansions and express them as polynomials in three invariants $\cal{F},\cal{G}$ and $\cal{H}$ \eqref{eq:invariants}.


\subsection{Spinor QED}

In general, the one-loop effective action of a given theory can be written as the functional trace of some differential operator $\Delta$. In the case of spinor electrodynamics, 
\bea\label{eq:effectiveactiondef}
    \Gamma^{(1)}[A]= -\ri \text{Tr}\ln \big(\ri \gamma^a \nabla_a-m\mathbbm{1}\big)=
    -\frac{\ri}{2}\text{Tr}\ln\Delta~, \qquad \Delta = -\square+m^2 \mathbbm{1}-\frac{\ri e}{2}\gamma^{ab}F_{ab}\,,
\eea
where $\square=\nabla^a\nabla_a$ is the covariant d'Alembertian. 
In what follows, our analysis will be restricted to the even-dimensional case, 
 $d=2n$.
 
 
 \subsection{Schwinger's formalism in $d$ dimensions}
 
 The effective action admits a representation in terms of the heat kernel $K(x,x';s)$ (see \cite{DeWitt:1964mxt,Barvinskii:2024iqz,Vassilevich:2003xt}),
\bea
    \Gamma^{(1)}[A] = -\frac{1}{2}\int_0^\infty \frac{\rd s}{\ri s}\int \rd^d x\,\tr K(x;s) ~, \qquad K(x;s)=\lim_{x'\to x}K(x,x';s)~.
\label{eq:effectiveaction}
\eea
The heat kernel associated with operator $\Delta$ is defined by\footnote{Here $\mathcal{I}(x,x')$ is the parallel displacement operator, and is supplied such that the heat-kernel transforms correctly under gauge transformations. It is normalised such that $\mathcal{I}(x,x) = \mathbbm{1}$.  See \cite{Kuzenko:2003eb} for a more in-depth treatment.}
\bea
    K(x,x';s)=\re^{-\ri\Delta s}\delta^{(d)}(x-x')\mathcal{I}(x,x')\,. \label{eq:heatkernel}
\eea
The  kernel obeys the equation 
\bea
    (\partial_s+\ri\Delta)K(x,x';s)=0~. \label{eq:definingrelation}
\eea
To calculate the one-loop effective action, we use Schwinger's operator approach \cite{Schwinger:1951nm}. Substituting momentum operators $\Pi_a =-\ri\nabla_a$ into \eqref{eq:effectiveactiondef} gives 
\bea
    \Delta = \Pi^a\Pi_a -\frac{\ri e}{2}\gamma^{ab}F_{ab}+m^2=-(\gamma^a\Pi_a)^2+m^2~.
\eea
We rewrite \eqref{eq:heatkernel} as
\bea
    K(x,x';s) =\re^{-\ri m^2 s}\braket{x|\re^{-\ri Hs}|x'}=\re^{-\ri m^2 s}\braket{x,s|x',0}~,
    \qquad H=-(\gamma^a\Pi_a)^2~,
\eea 
where $H$ takes the role of the generator of time evolution with respect to the `proper-time' parameter $s$. There are also three important relations that characterise the heat kernel,
\begin{subequations}
\begin{align}
    \ri\partial_s\braket{x,s|x',0} &= \braket{x,s|H|x',0}, \label{eq:eq1}\\
    \braket{x,s|\Pi_a(s)|x',0} &= -\ri\nabla_a\braket{x,s|x',0} ,\label{eq:eq2}\\
    \braket{x,s|\Pi_a(0)|x',0} &= \ri\nabla'_a\braket{x,s|x',0} .\label{eq:eq3}
\end{align}
\end{subequations}
The Heisenberg equations read
\begin{subequations}
\begin{align}
    \frac{\rd{x^a}}{\rd s} &= -\ri[x^a,H]~,\\
    \frac{\rd{\Pi_a}}{\rd s} &= -\ri[\Pi_a,H]~.
\end{align}
\end{subequations}
Solving the Heisenberg equations allows us to write $H$ in terms of the operators $x^a(s)$ and $x'^a(0)$. This allows us to solve \eqref{eq:eq1} for the heat kernel. For a constant field in even-$d$ dimensions the technical details of the calculation are much the same as in \cite{Schwinger:1951nm,Dittrich:1985yb,Dunne:2004nc}. The final result is
\begin{multline}\label{eq:heatkernelclosed}
    \braket{x,s|x',0} = \frac{\ri}{(4\pi \ri s)^{d/2}}\sqrt{\det\left(\frac{e\bm{F}s}{\sinh (e\bm{F}s)}\right)}\times\\\times\exp\left(\frac{\ri}{4}(x-x')e\bm{F}\coth{(e\bm{F} s)}(x-x')\right)\exp\left(-\frac{1}{2}es\gamma^{ab}F_{ab} \right)\,,
\end{multline}
where we have introduced the matrix
\bea
    \bm{F} = ({F^a}_b)\,.
    \label{fatF}
\eea
The effective action is obtained by inserting the coincidence limit of \eqref{eq:heatkernelclosed} into \eqref{eq:effectiveaction}. To ensure convergence of the proper-time integral for large $s$,
we perform a Wick rotation $s \to -\ri s$. We obtain the effective Lagrangian,
\bea
    \mathcal{L}^{(1)} = -\frac{1}{2}\int_0^\infty \frac{\rd s}{s}\frac{1}{(4\pi s)^{d/2}}\re^{-m^2 s}\sqrt{\det\left(\frac{e\bm{F}s}{\sin (e\bm{F}s)}\right)}\tr\exp\left(\frac{\ri }{2} es
    \gamma^{ab}F_{ab}\right). \label{eq:action1}
\eea
The above is the generic result in $d=2n$ dimensions. In this form, the Lagrangian is difficult to work with and the computation of weak-field limits in terms of invariants of the electromagnetic field is nontrivial. In what follows, we provide an approach to do so, first in six dimensions and then in general.


\subsection{6D Spinor QED} 

Let us first specialise to the $d=6$ case. We evaluate the spinor trace appearing in \eqref{eq:action1}. We decompose $\gamma^{ab}$ into left and right-handed Weyl spinor representations \eqref{Lorentz} to obtain the relation
\begin{align}
    \tr \exp\left(\frac{\ri}{2}es\gamma^{ab}F_{ab} \right) &= \tr\exp\left(\frac{\ri}{2}es\sigma^{ab}F_{ab} \right) +\tr\exp\left( -\frac{\ri}{2}es\sigma^{ab}F_{ab}\right)~, \non\\
    &=2\tr\cos\left(\frac{1}{2}es\sigma^{ab}F_{ab}\right)~.
    \label{trace2.14}
\end{align}
In the exponent, by eq. \eqref{eq:correspondence2} we have the $4\times 4$ matrix
\bea
      \mathbb{F} = ({F_{\alpha}}^\beta )= - \frac 14  \sigma^{ab}F_{ab} ~,
\eea
and so \eqref{trace2.14} becomes
\bea
    \tr \exp\left(\frac{\ri}{2}es\gamma^{ab}F_{ab} \right) = 2\tr\cos\left(2es\mathbb{F}\right)~.
      \label{trace2.16}
\eea

We obtain the eigenvalues of $\mathbb{F}$ using the 6D spinor formalism. In \eqref{eq:spinorcharacteristic}, we show that they are given by solutions to the quartic equation
\bea
    \omega^4-\mathcal{F}\omega^2-\mathcal{G}\omega+\frac{1}{4}(\mathcal{H}-\mathcal{F}^2)=0~. \label{eq:eigenvalue1}
\eea
Here the coefficients of the quartic polynomial are expressed in terms of the invariants of the electromagnetic field introduced in Appendix \ref{appendixB}.
Similarly, in \eqref{eq:tensorcharacteristic} we show that the eigenvalues of $\bm{F}$ are given by solutions to the equation
\bea
    \lambda^6-2\mathcal{F}\lambda^4+(2\mathcal{F}^2-\mathcal{H})\lambda^2 -\mathcal{G}^2=0~.
    \label{eq:eigenvalue2}
\eea

Let us make the following ans\"atze:
\begin{subequations}
    \begin{align}
        \lambda_1^2 &= (\omega_1+\omega_2)^2~,\\
        \lambda_2^2 &= (\omega_1+\omega_3)^2~,\\
        \lambda_2^3 &= (\omega_2+\omega_3)^2~.
    \end{align}
\end{subequations}
Using Vi\`ete's formulas (see, e.g., Section 3.2 in \cite{Vinberg}) 
one can show that the $\omega_i$ solve \eqref{eq:eigenvalue1} if and only if the $\lambda_i$ solve \eqref{eq:eigenvalue2}. This means that the roots of the above polynomials for the eigenvalues of $\bm{F}$ and $\mathbb{F}$ satisfy the following relations\footnote{It may be shown that the eigenvalues of 
$ \tilde{\mathbb{F}} = - \frac14  \tilde{\sigma}^{ab}F_{ab} $ are $\tilde{\omega}_i = -\omega_i$, with $i=1,2,3,4$.}
\begin{subequations}  \label{eq:wlrelation}
\begin{align}
    \omega_{1,2} &= \frac{1}{2}\lambda_1\pm\frac{1}{2}(\lambda_2-\lambda_3)~,\\
    \omega_{3,4} &= -\frac{1}{2}\lambda_1\pm\frac{1}{2}(\lambda_2+\lambda_3)~.
\end{align}
\end{subequations}
We can make use of \eqref{eq:wlrelation} to represent \eqref{trace2.16}  as
\begin{subequations} 
    \begin{align}
    \tr \exp\left(\frac{\ri}{2}es\gamma^{ab}F_{ab} \right) &= \sum_{i=1}^4 2\cos(2es\omega_{i})\label{eigenvaluesum}\\
    &= 8\cos(es\lambda_1)\cos(es\lambda_2)\cos(es\lambda_3)\label{eigenvalueprod}~,
\end{align}
\end{subequations}
where we have used 
standard trigonometric identities to convert the sum in \eqref{eigenvaluesum} to the product in \eqref{eigenvalueprod}. 
We also have
\begin{align} 
    \sqrt{\det\left(\frac{e\bm{F}s}{\sin(e\bm{F}s)}\right)}
    &=  \frac{(es)^3\lambda_1 \lambda_2\lambda_3}{\sin(es\lambda_1)\sin(es\lambda_2)\sin(es\lambda_3)}~,
\end{align}
and so the renormalised effective Lagrangian is
\begin{align}
    \mathcal{L}_{\text{spinor}}^{(1)} = 
    -\frac{1}{2}\int_0^\infty \frac{\rd s}{s}\frac{8}{(4\pi s)^3}\re^{-m^2 s}[(es)^3\lambda_1 \lambda_2 \lambda_3\cot(es\lambda_1)\cot(es\lambda_2)\cot(es\lambda_3)-1+\frac{2}{3}(es)^2\mathcal{F}]~,\label{6DEH}\end{align}
where the appropriate counterterms have been supplied to renormalise the vacuum energy and Maxwell term. 

We now turn to the weak-field expansion. The weak field limit corresponds to small $\lambda_1,\lambda_2,\lambda_3$, and so we can expand the cotangent product in the integrand. Since it is symmetric in the eigenvalues, its series expansion must consist of symmetric polynomials in the eigenvalues. The fundamental theorem of symmetric polynomials (see, e.g., Section 3.8 in \cite{Vinberg}) thus allows us to express the weak-field expansion in powers of 
the invariants $\mathcal{F},\mathcal{G}$ and $\mathcal{H}$, which are defined in  \eqref{eq:invariants} and are 
expressed in terms of the eigenvalues $\lambda_i $ according to eq. \eqref{eq:vieta}. The resulting expansion reads
\begin{align}
     \mathcal{L}^{(1)}_{\text{spinor}}=
     -\frac{e^4}{720\pi^3 m^2}(10\mathcal{F}^2-7\mathcal{H})-\frac{e^6}{3780\pi^3 m^6}(18\mathcal{F}^3-13\mathcal{F}\mathcal{H}-31\mathcal{G}^2)+...
\end{align}


\subsection{6D Scalar QED}
As demonstrated in \cite{Dittrich:1985yb}, the one-loop effective action for scalar QED is given by,
\bea
    \Gamma^{(1)}[A]=\ri\text{Tr}\ln\Delta,\quad \Delta = -\square+m^2~.\label{scalareffectiveaction}
\eea
This is identical in form to \eqref{eq:effectiveactiondef}, except multiplied by a factor of $-2$ and omitting the spinor trace in the heat kernel. We can immediately write down the effective Lagrangian, 
\begin{align}
    \mathcal{L}^{(1)}_{\text{scalar}} = 
    \int_0^\infty \frac{\rd s}{s}\frac{1}{(4\pi s)^3}\re^{-m^2 s}[(es)^3\lambda_1 \lambda_2 \lambda_3\csc(es\lambda_1)\csc(es\lambda_2)\csc(es\lambda_3)-1-\frac{1}{3}(es)^2\mathcal{F}]~, \label{eq:scalarcase}
\end{align}
where again the appropriate counterterms have been supplied for renormalisation.

Performing a similar weak-field expansion on this action and collecting the symmetric polynomials in the eigenvalues gives the expansion in terms of the invariants $\mathcal{F}$, $\mathcal{G}$, $\mathcal{H}$,
\begin{align}\label{quarticwf}
    \mathcal{L}^{(1)}_{\text{scalar}} = 
    \frac{e^4}{5760\pi^3 m^2}(5\mathcal{F}^2+\mathcal{H})+\frac{e^6}{60480\pi^3 m^6}(9\mathcal{F}^3+11\mathcal{F}\mathcal{H}+2\mathcal{G}^2)+...
\end{align}


\section{Explicit $d$-dimensional results}\label{section3}

As an extension of the $d=6$ analysis in the previous section, here we provide an explicit computation of the Euler-Heisenberg Lagrangian in $d=2n>4$ 
dimensions, for both spinor and scalar QED, in terms of electromagnetic invariants.

Our task includes expressing  
\bea
\tr \exp\left(\frac{\ri}{2}es\g^{ab}F_{ab}\right)
\eea
in terms of the electromagnetic invariants 
 beyond four dimensions. Schwinger's original $d=4$ calculation made use of the identity
\bea
\left(\frac{\ri}{2}\g_{ab}F^{ab}\right)^2=\frac{1}{2}F_{ab}F^{ab}{\mathbbm 1}+\frac{1}{2} F_{ab}\tilde{F}^{ab}\gamma_5~,
\qquad \widetilde{F}^{ab} = \hf \ve^{abcd} F_{cd}~.
\eea
Here the right-hand side involves the two invariants of the electromagnetic field, 
$\mathcal{S} =  -\frac 14 F^{ab} F_{ab}$ and $\mathcal{P} = \frac 14 \widetilde{F}^{ab} F_{ab}$.
This approach requires a generalisation to work in higher dimensions. In $d=2n$ dimensions there are $n$ independent invariants, so finding a similar expression in general is not easy. One could, in principle, use the spinor technology developed in Appendix \ref{appendixA} to perform a similar analysis to the above six-dimensional case, but since the Fierz identities for $\s_{ab}$ depend on the spacetime dimension we also anticipate that generalising the result to arbitrary $d$ using this approach would be challenging.

Another complication comes from a difference between the Euclidean and Minkowski spaces. 
In Euclidean space, ${\mathbb E}^{2n}$, every real antisymmetric tensor $F=(F_{ab})$ can be brought to a canonical 
block-diagonal form by a rotation,
\bea
O FO^{\rm T} 
= \bigoplus_{k=1}^{d/2}\begin{pmatrix}
    0 & f_k\\
    -f_k & 0
\end{pmatrix}~, 
\qquad O \in \sSO(2n)~.
\label{CanForm}
\eea
This form is useful for computing Euler-Heisenberg-like Lagrangians, in particular for evaluating the structure
$    \det \big( {e\bm{F}s} \big[\sin (e\bm{F}s)\big]^{-1}\big)$
    in \eqref{eq:action1}.

 In Minkowski space, ${\mathbb M}^{2n}$, 
the situation is more involved. For simplicity, let us consider the $d=4$ case where the electromagnetic field strength is
\bea
F = \begin{pmatrix}
    0 & -E_x&-E_y &-E_z\\
    E_x &0&B_z&-B_y\\
    E_y&-B_z&0&B_x\\
    E_z & B_y&-B_x &0
\end{pmatrix}~.
\eea
There are two special cases, see, e.g., \cite{Rainich:1925qbz} and \cite{LL} (section 24).
\begin{itemize} 
\item
If at least one of the invariants  $\mathcal{S}$ and $\mathcal{P}$ is non-vanishing,  an inertial system exists in which $\vec E$ and $\vec B$ are parallel. One may find a Lorentz transformation $\Lambda$ such that 
\bea
\Lambda F \Lambda^{\rm{T}}  = \begin{pmatrix}
    0 & -E&0 &0\\
    E &0&0&0\\
    0&0&0&B\\
    0 & 0&-B &0
\end{pmatrix}~,\qquad \Lambda \in \sSO_0(3,1)~,\label{EBCanForm}
\eea
which is of the type \eqref{CanForm}.

\item 
If both invariants the electromagnetic field vanish, $\mathcal{S}=\mathcal{P} =0$, it is not possible to bring $F_{ab}$ to a block-diagonal form. Such $F_{ab}$ is referred to as a {\it null} field, see e.g.  \cite{Misner:1957mt}. Algebraically, $F_{ab}$ is generated by a null vector $p^a$ and a space-like vector $v^a$, 
\bea
F_{ab} = p_a v_b - p_b v_a~, \qquad p^a p_a =0~, \quad p^a v_a =0~, \quad v^a v_a > 1~.
\label{null}
\eea
One may perform  a Lorentz transformation $ \Lambda \in \sSO_0(3,1)$ such that $F$ takes the form 
\bea
F' :=\Lambda F \Lambda^{\rm T}   = \begin{pmatrix}
    0 & 0&-E &0\\
    0 &0&E&0\\
    E&-E&0&0\\
    0 & 0&0 &0
\end{pmatrix}\quad \implies \quad (\bm{F}')^3 =0~,
\eea
with $\bm{F} $ defined in \eqref{fatF}. 
\end{itemize}
We conclude that in Minkowski signature, it is not always possible to  block diagonalise the field strength. In what follows, we will  assume that $\bm{F}$ is not null, and as such the canonical form \eqref{EBCanForm} is of use. Since $\bm{F}$ is a skew-symmetric endomorphism, using the terminology of  \cite{Mars:2020yer}, it follows that for any non-null field there is an orthonormal basis in which $\bm{F}$ decomposes into a direct sum of the $\mathbb{M}^{4}$ canonical form \eqref{EBCanForm} and $2\times2$ spacelike eigenplanes $\bm{F}\vert_{\Pi_i}$,
\bea
\bm{F}= \bm{F}\vert_{\mathbb{M}^{4}}\bigoplus_{i=1}^{n-2}\bm{F}\vert_{\Pi_i}~.
\eea
In an appropriate orthonormal basis, the $2\times 2$ blocks $\bm{F}\vert_{\Pi_i}$ take the form
\bea
\bm{F}\vert_{\Pi_i}= \begin{pmatrix}
    0 & \mu_i\\
    -\mu_i &0
\end{pmatrix}~.
\eea
Its eigenvalues are $\l_i = \pm\ri\mu_i$.

We now express the effective action in terms of the eigenvalues of $\bm{F}$. 
The eigenvalues of the four-dimensional matrix $\bm{F}\vert_{\mathbb{M}^{4}}$, which corresponds to the 
canonical form \eqref{EBCanForm}, are $\pm E$ and $\pm \ri B$. Since the eigenvalues of an endomorphism are invariant under Lorentz transformations, we may identify
\bea
E^2 = \l_1^2~, \quad B^2 = -\l_2^2~,
\eea
where
\bea
\l_1^2 &= \mathcal{S} + \sqrt{\mathcal{S}^2+\mathcal{P}^2}~,\quad \l_2^2 =  \mathcal{S} - \sqrt{\mathcal{S}^2+\mathcal{P}^2}~.\label{squareEVs}
\eea
Note that $\l_1$ is always real and $\l_2$ is always imaginary. In terms of these eigenvalues,
\bea
\bm{F}\vert_{\mathbb{M}^{4}} = \begin{pmatrix}
    0 & \l_1&0 &0\\
    \l_1 &0&0&0\\
    0&0&0&\ri \l_2\\
    0 & 0&-\ri\l_2 &0
\end{pmatrix}~.\label{canformeig}
\eea
The signs of the eigenvalues in \eqref{canformeig} are arbitrary and may be chosen without loss of generality. Indeed, the effective action is invariant under the change $\l_i \to -\l_i$. This is because all invariants are functions only of $\l_i^2$.

We find that there is a basis in which, for non-null fields, it is possible to bring the field strength to the canonical form
\bea
\bm{F} = \begin{pmatrix}
    0 &\l_1 \\
    \l_1 & 0
\end{pmatrix} \bigoplus_{i=2}^{n}\begin{pmatrix}
    0 & \ri\l_i\\
    -\ri\l_i &0
\end{pmatrix}~.
\eea
Here now $\l_i$ (including $\l_1$) are eigenvalues of the $d$-dimensional $\bm{F}$, as the eigenvalues of a skew-symmetric matrix do not change under direct summation with $2\times 2$ off-diagonal blocks. 

We proceed now by computing $\g^{ab}F_{ab}$. The decomposition of $F_{ab}$ into canonical form allows this expression to fully factor as,
\bea
\g^{ab}F_{ab} = -2\g^0\g^1\l_1+2\ri\g^2\g^3\l_2 +\dots + 2\ri\g^{d-2}\g^{d-1}\l_{d/2}~.\label{gmabapp}
\eea
We also note that
\bea
    (\g^i\g^{i+1})^2 = \begin{cases} 1~, \,\quad i=0\\
    -1~, \:\,i\geq2
\end{cases}\label{eigenvalueresult}
\eea
The family of matrices of the form $\g^i\g^{i+1}$ with disjoint index pairs are normal and mutually commute, and as such share a common basis of eigenvectors. Acting with the expression \eqref{gmabapp} on one of these shared eigenvectors $\bm{v}$ gives,
\bea
\g^{ab}F_{ab}\bm{v} = (2\ve_1 \l_1 + 2\ve_2 \l_2 +\dots + 2\ve_{d/2}\l_{d/2})\bm{v}~,
\eea
where, by \eqref{eigenvalueresult}, each $\ve_i$ is some choice of $\pm 1$. Different eigenvectors $\bm{v}$ realise all possible sign combinations. To see this, one can notice that it is simple to generate new eigenvectors by multiplying $\bm{v}$ by some $\g^a$, and that each of these new eigenvectors correspond to a different sign choice for each $\ve_i$. We find that the eigenvalues of the matrix $\g^{ab}F_{ab}$ are,
\bea
\sum_{i=1}^{d/2}2\ve_i\l_i~, \quad \ve_i = \pm 1~,
\eea
where each $\ve_i$ is chosen freely. Now,
\bea
\tr \exp\left(\frac{\ri}{2}es\g^{ab}F_{ab} \right) &=& \sum_{\ve_i =  \pm 1}\exp\left(\ri e s \sum_{i=1}^{d/2} \ve_i \l_i \right)= 2^{d/2}\prod_{i=1}^{d/2} \cos(es\l_i)~,
\eea
and
\bea
\sqrt{\det\left(\frac{e\bm{F}s}{\sin{e\bm{F}s}}\right)} = (es)^{d/2}\prod_{i=1}^{d/2}\frac{\l_i}{\sin(es\l_i)}~.
\eea

The final expression for the Euler-Heisenberg action is then,
\bea 
\mathcal{L}_{\text{spinor}}^{(1)} = -\frac{1}{2}\int_0^\infty \frac{\text{d}s}{s}\frac{2^{d/2}}{(4\pi s)^{d/2}}\re^{-m^2 s}(es)^{d/2}\prod_{i=1}^{d/2}\l_i \cot(es\l_i)~.\label{ddim}
\eea
The scalar QED case follows trivially; we multiply by $-2$ and omit the contribution from the trace,
\bea 
\mathcal{L}_{\text{scalar}}^{(1)} = \int_0^\infty \frac{\text{d}s}{s}\frac{1}{(4\pi s)^{d/2}}\re^{-m^2 s}(es)^{d/2}\prod_{i=1}^{d/2}\l_i \csc(es\l_i)~.\label{ddimsc}
\eea

One may choose a set of $n=d/2$ invariants, which are most commonly proportional to traces of even powers of the field strength, 
$\tr(\bm{F}^{2k})$, 
along with a Pfaffian-type invariant. The independent invariants\footnote{All the invariants vanish in the null case \eqref{null}.} 
can be chosen as follows:
\begin{subequations} \label{Invariants}
\bea
\mathcal{S}_k&=&\frac{1}{4}\tr(\bm{F}^{2k})~,\qquad 1 \leq k \leq n-1~, \\
 \mathcal{P} &=& \frac{1}{2^n n!}\ve^{a_1 a_2...a_d} F_{a_1 a_2}F_{a_3 a_4}\dots F_{a_{d-1}a_d}~, \quad \det(\bm{F}) = -\mathcal{P}^2~.
\eea
\end{subequations}
It is an instructive exercise to express
invariants $\mathcal{S}_k=\frac{1}{4}\tr(\bm{F}^{2k})$, with $k\geq n$, 
in terms of \eqref{Invariants}.
The characteristic equation of $\bm{F}$ in $d=2n$ dimensions is
\begin{align}
\l^d + e_2 \l^{d-2} + e_4 \l^{d-4} +\dots -\mathcal{P}^2=0 ~ \implies ~
{\bm F}^d + e_2 {\bm F}^{d-2} + e_4 {\bm F}^{d-4} +\dots -\mathcal{P}^2 {\mathbbm 1}_d=0 ~,
\end{align}
where the even elementary symmetric polynomials $e_{2n}$ can be written in terms of the invariants $\mathcal{S}_k$ in the following way,
\bea\label{ESP}
e_{2n} &=& -\frac{2}{n}\sum_{i=0}^{n-1}e_{2i} \mathcal{S}_{n-i}~, \qquad e_0 =1~.
\eea
Viet\`e's formulas then allow one to write the series expansion of the above Lagrangian directly in terms of one's choice of invariants. 
\begin{subequations}
\bea
\sum_{i=1}^{n}\l_i^2 &=& -e_2~,\\
\sum_{1\leq i < j \leq n}\l_i^2\l_j^2 &=& e_4~,\\
\sum_{1\leq i < j<k \leq n}\l_i^2\l_j^2\l_k^2 &=& -e_6~,\\
&\vdots\notag\\
\l_1^2\l_2^2...\l_{n}^2 &=& -(-1)^{n}\mathcal{P}^2~.
\eea
\end{subequations}

In this way, it is possible to write the symmetric polynomials in the eigenvalues that appear in the weak field expansion of the effective action directly in terms of $\mathcal{S}_n$. As an application of our approach, we compute the finite corrections to the classical Lagrangian for spinor and scalar QED in 8 and 10 dimensions.
\bea
\mathcal{L}^{(8\text{D})}_{\text{spinor}} &=& \frac{e^6}{45360 m^4 \pi^4}(70\mathcal{S}_1^3-147\mathcal{S}_1\mathcal{S}_2+62\mathcal{S}_3)\notag\\
&+&\frac{e^8}{226800m^8\pi^4}(412 \mathcal{S}_1^4 + 1143 \mathcal{P}^2 - 816 \mathcal{S}_1^2 \mathcal{S}_2 + 57 \mathcal{S}_2^2 + 284\mathcal{S}_1 \mathcal{S}_3)+ ...~,\\
\mathcal{L}^{(8\text{D})}_{\text{scalar}} &=& \frac{e^6}{1451520m^4 \pi^4}(35\mathcal{S}_1^3+21\mathcal{S}_1\mathcal{S}_2+4\mathcal{S}_3)\notag\\
&+&\frac{e^8}{14515200m^8\pi^4}(199\mathcal{S}_1^4+36\mathcal{P}^2+138\mathcal{S}_1^2\mathcal{S}_2+39\mathcal{S}_2^2+128\mathcal{S}_1\mathcal{S}_3)+ ...~.\\
\mathcal{L}^{(10\text{D})}_{\text{spinor}} &=& \frac{e^6}{90720m^2\pi^5}(70\mathcal{S}_1^3-147\mathcal{S}_1\mathcal{S}_2+62\mathcal{S}_3)\notag\\
&-&\frac{e^8}{2721600m^6\pi^5}(700\mathcal{S}_1^4-2940\mathcal{S}_1^2\mathcal{S}_2+1029\mathcal{S}_2^2+2480\mathcal{S}_1\mathcal{S}_3-1143\mathcal{S}_4)+...~,\\
\mathcal{L}^{(10\text{D})}_{\text{scalar}} &=& \frac{e^6}{5806080m^2\pi^5}(35\mathcal{S}_1^3+21\mathcal{S}_1\mathcal{S}_2+4\mathcal{S}_3)\notag\\
&+&\frac{e^8}{174182400m^6\pi^5}(175\mathcal{S}_1^4+210\mathcal{S}_1^2\mathcal{S}_2+21\mathcal{S}_2^2+80\mathcal{S}_1\mathcal{S}_3+18\mathcal{S}_4)+...~.
\eea
In the case of null fields, where all invariants vanish, 
these corrections vanish.


\section{Pair production in $d$ dimensions}\label{section4}

In this section we compute the rate of pair production in a background electric field in both spinor and scalar QED in $d$ dimensions.
The pair production rate is obtained from the vacuum persistence amplitude \cite{Schwinger:1951nm}. To one-loop order, it is given by
\bea
    \braket{0_{\text{out}}|0_{\text{in}}} = Z[A] =\re^{\ri\Gamma^{(1)}[A]}~. \label{eq:vacuumpersistence}
\eea
From \eqref{eq:vacuumpersistence}, to first order we have
\begin{align}
    |{\braket{0_{\text{out}}|0_{\text{in}}}}|^2 &= \re^{\ri\Gamma^{(1)}[A]}\re^{-\ri{\bar{\Gamma}}^{(1)}[A]}\\
    &=\re^{-2\text{Im}\,\Gamma^{(1)}[A]}\\
    &\approx 1-2\text{Im}\,\Gamma^{(1)}[A]~.
\end{align}
The probability per unit time per unit volume that a pair is produced is
\bea
    p=2\text{Im}\,\mathcal{L}^{(1)}~.
\eea
\subsection{Spinor QED}
We wish to consider the Euler-Heisenberg Lagrangian in a constant electric field and no magnetic field. The field strength reads
\bea
    F_{i0} = E_i =-F_{0i}~,\quad F_{ij}=0~.
\eea
In this configuration, the eigenvalues of the field strength are simply,
\bea
\pm \l = \pm E~.
\eea
Substituting this into the Lagrangian \eqref{ddim} yields,
\bea
    \mathcal{L} = \frac{-2^{d/2} eE}{2(4\pi)^{d/2}}\int_0^\infty \frac{\rd s}{s^{d/2}}\re^{-m^2 s}\cot{(eEs)}\,.
\eea
The imaginary part of the Lagrangian comes from the poles of the cotangent at $s_n = \frac{n\pi}{eE}$. Evaluating the integral by deforming the contour around the poles in the upper-half plane gives the contribution to the pair production rate in $d$ dimensions,
\begin{align}
    p=2{\rm Im}\,\mathcal{L} &= \frac{-2^{d/2}}{(4\pi)^{d/2}}\left(-\pi \sum_{n=1}^{\infty} \frac{(e E)^{d/2}}{(n\pi)^{d/2}}\re^\frac{-nm^2\pi}{eE}\right)
    =\frac{2^{d/2}\pi(eE)^{d/2}}{(4\pi)^{d/2}}\sum_{n=1}^{\infty} \frac{1}{(n\pi)^{d/2}}\re^\frac{-nm^2\pi}{eE}\,.
\end{align}
Substituting $d=4$ yields,
\bea
    p=\frac{(eE)^2}{4\pi^3}\sum_{n=1}^{\infty}\frac{1}{n^2}\re^\frac{-nm^2\pi}{eE}\,.
\eea
which is the known result first derived by Schwinger \cite{Schwinger:1951nm}.

\subsection{Scalar QED}
For the case of scalar QED, the result is essentially identical, except we omit the factor of $2^{d/2}$ that comes from the trace over $\gamma^{ab}$. Also, the sign of the residue flips at every pole of $\csc(eEs)$, introducing a factor of $(-1)^{n+1}$. Considering the extra factor of $2$ in \eqref{ddimsc}, we find that the pair production rate is
\bea
    p =\frac{2\pi(eE)^{d/2}}{(4\pi)^{d/2}}\sum_{n=1}^{\infty} \frac{(-1)^{n+1}}{(n\pi)^{d/2}}\re^\frac{-nm^2\pi}{eE}\,.
\eea


\section{Conformal actions in six and eight dimensions}\label{section5}

In this section we discuss conformal actions for the electromagnetic field in $d=2n >4$  dimensions. 
Such actions naturally occur as logarithmic divergences of the one-loop effective actions for massless scalar and spinor fields coupled to the electromagnetic field. 
There is a unique conformal action of second order in the field strength
\bea
\int \rd^{d}  x\,F^{ab}\square^{(d-4)/2} F_{ab} ~.
\label{6d-action}
\eea
In eight and higher dimensions, conformal actions exist that involve higher powers of the field strength.
An example is provided by an eight-dimensional action of the form
\begin{align} 
 \int \rd^8 x\, \Big\{ \alpha \,{\rm tr} \big( {\bm F}^4 \big) +\beta \,{\rm tr} \big( {\bm F}^2 \big) {\rm tr} \big( {\bm F}^2 \big) 
\Big\}~,
\label{8d-action-b}
\end{align}
for some numerical coefficients $\alpha$ and $\beta$. 
Below we elaborate on the conformal properties of these functionals. 

Given a conformal Killing vector field $\xi = \xi^a \pa_a$ on ${\mathbb M}^d$, 
\bea
\pa_a \xi_b + \pa_b \xi_a = \frac{2}{d} \eta_{ab} \pa_c \xi^c~,
\label{CKE}
\eea
there are two building blocks, 
\bea
K_{ab}[\xi] = \hf (\pa_a \xi_b - \pa_b \xi_a )~, \qquad \sigma[\xi] = \frac{1}{d} \pa_c \xi^c~,
\label{blocks}
\eea
which determine the conformal transformation laws of conformal primary fields.
The transformation law of a conformal primary field $\Phi$ (with suppressed indices) of dimension $\Delta$ is 
\bea
\d_\xi \Phi = \Big( \xi +\hf K^{bc} M_{bc}  + \Delta \sigma[\xi]  \Big) \Phi
\equiv  \Big( \cK_\xi  + \Delta \sigma[\xi]  \Big) \Phi
~,
\eea
with $M_{bc}$ the Lorentz generator.
In particular, the conformal transformation law of the electromagnetic field strength $F_{ab}$ is
\bea
\d_\x F_{ab} = \Big( \xi + 2 \sigma[\xi]\Big) +K_a{}^c[\xi] F_{cb} + K_b{}^c [\xi] F_{ac}~.
\label{F-CTL}
\eea
Given a conformal primary scalar field $\cL$ of dimension $d$, the functional 
\bea
\int \rd^d x\,\cL
\eea
is conformally invariant. Since ${\rm tr} \big( {\bm F}^2 \big)$ and 
${\rm tr} \big( {\bm F}^4 \big)$ are conformal primary fields of dimension $+4$ and $+8$, respectively,
the conformal invariance of \eqref{8d-action-b} follows.

In order to prove conformal invariance of the functional \eqref{6d-action}, it is helpful to make use of the identities 
\bea
\pa_a K_{bc}[\xi] = \eta_{ac} \pa_b \sigma[\xi] - \eta_{ab} \pa_c \sigma[\xi]~, \qquad 
\pa_a \pa_b \sigma[\xi] =0~,
\eea
which follow from the conformal Killing equation \eqref{CKE}, or its equivalent form 
\bea
\big[ \xi + \hf K^{bc} M_{bc} \, , \, \pa_a\big] + \s \pa_a - \pa^b \s M_{ab}=0~.
\eea 
Then one may deduce from \eqref{F-CTL}
\begin{align}
\square^m \d_\xi F_{ab} =& \Big( \cK_\xi  + (2m+2) \sigma[\xi]  \Big) \square^m F_{ab}
+2m (m+1 -d/2) \pa^c\sigma[\xi] \pa_c \square^{m-1} F_{ab} \non \\
& +2m \Big( \pa_b \sigma[\xi] \square^{m-1}\pa^c F_{ac} - \pa_a \sigma[\xi] \square^{m-1}\pa^c F_{bc} \Big)~.
\end{align}

In Appendix \ref{appendixC} we evaluated the $a_3$ coefficients, eqs. \eqref{eq:traces} and \eqref{a3-scalar}, which 
determine the one-loop logarithmic divergences in spinor and scalar QED.
The  logarithmic divergence is proportional to the conformally invariant functional 
\bea
\int \rd^6 x\,F^{ab}\square F_{ab}~.
\eea
In curved space, the logarithmic divergent contribution to the effective action should have the form 
\bea
\int \rd^6 x\, e\,  I ~+~ \text{total derivative} ~, \qquad e=\det (e_m{}^a)~,
\eea
where $I$ is a conformal primary field of dimension $6$ constructed from the field strength $F_{ab}$.  In $d=2n \neq 4$ dimensions, there is a unique primary field with the required properties. Following the formalism of conformal gravity 
described in Section 2 of \cite{Butter:2016qkx} (see also Section 3 of \cite{BKNT-M1} and Section 2 in \cite{Kuzenko:2026jwd}), 
the required composite primary field is
\bea
I = F^{bc} \Box_{\rm c} F_{bc} + \frac{10-d}{3(d-4)} {\bm \nabla}_c F^{bc} {\bm \nabla}^d F_{bd} 
+\frac{d-4}{6} {\bm \nabla}^d F^{bc} {\bm \nabla}_dF_{bc}~, \qquad \Box_{\rm c} = {\bm \nabla}^d {\bm \nabla}_d~,
\label{primary}
\eea
Here ${\bm \nabla}_a$ is the conformally covariant derivative. The scalar $I$ is conformal primary in the sense that it is annihilated by the special conformal generator, 
\bea
K_a I =0~.
\eea
For the three structures in \eqref{primary} one finds
\begin{subequations}
\bea
K_a  F^{bc} \Box_{\rm c} F_{bc} &=& 2(4-d) F^{bc}{\bm \nabla}_a F_{bc} +8 F_{ba} {\bm \nabla}_c F^{bc}~,\\
K_a {\bm \nabla}_c F^{bc} {\bm \nabla}^d F_{bd} &=& 4 (4-d) F_{ba} {\bm \nabla}_c F^{bc} ~, \\
K_a  {\bm \nabla}^d F^{bc} {\bm \nabla}_dF_{bc}&=& 12 F^{bc}{\bm \nabla}_a F_{bc} - 8 F_{ba} {\bm \nabla}_c F^{bc}~.
\eea
\end{subequations}
The second term in \eqref{primary} is ill-defined in the $d=4$ case since ${\bm \nabla}_b F^{ab}$ is a primary field in four dimensions. 
Primary field $(d-4)I $ is well-defined for any $d$. It is worth pointing out that making use of the Bianchi identity ${\bm \nabla}_{[a} F_{bc]} =0$ leads to the identity
\bea
\Box_{\rm c} F_{ab}+{\bm \nabla}_a {\bm \nabla}^c F_{bc} - {\bm \nabla}_b {\bm \nabla}^c F_{ac}
+C_{abcd} F^{cd}=0~,
\eea
where $C_{abcd}$ is the Weyl tensor. This identity allows one to rewrite the primary field \eqref{primary} in a different 
form.

Upon degauging (see \cite{Butter:2016qkx, BKNT-M1} for the technical details), the invariant  \eqref{primary}
takes the form 
\bea
I = F^{bc} \Big( \cD^d \cD_d +\frac{R}{d-1}\Big)  F_{bc} + \frac{10-d}{3(d-4)} \cD_c F^{bc} \cD^d F_{bd} 
+\frac{d-4}{6} \cD^d F^{bc} \cD_dF_{bc}~, 
\label{primary2}
\eea
where $\cD_a$ is the torsion-free Lorentz covariant derivative, and $R$ is the scalar curvature.  
This conformal scalar of dimension $+6$ was introduced in \cite{ParkerRosenberg} (see also \cite{Osborn:2015rna}).

In general, given a classically Weyl-invariant theory in $d=2n$ dimensions, 
the Weyl anomaly has the general form (see \cite{Duff:1993wm, BPB, DS, KMM, Bastianelli:2000hi, Casarin:2024qdn}
and references therein)
\bea
\langle T \rangle  = a \mathbb{E}_d + \mathfrak{I}  + \cD_a J^a~,
\label{anomaly}
\eea
where $\mathbb{E}_d$ is the Euler density, and $\mathfrak I$ is a conformal primary field of dimension $d$.
The second term in \eqref{anomaly} includes contributions from the gravitational field (composite fields constructed from the Weyl tensor and its covariant derivatives) and from background matter fields such as \eqref{primary2} in six dimensions. The third term in \eqref{anomaly} is generated by a local counterterm and, therefore, can be removed. 
Using the electromagnetic field strength, in the $d=6$ case one can construct two local counterterms
 of dimensions $+6$ that are quadratic in $F_{ab}$,
\bea
R F^{ab} F_{ab}~, \qquad R^{ab} F_{ac} F_{b d} \eta^{cd}~,
\eea
and contribute to the total derivative in \eqref{anomaly}. One more structure of dimensions $+6$ is
\bea
F^{ab} C_{abcd} F^{cd}~.
\label{primary3}
\eea
It generates a Weyl-invariant functional in six dimensions.

Am interesting open problem is to find an eight-dimensional analogue  of \eqref{primary}, that is a conformal primary completion of 
\bea
F^{bc} \Box_{\rm c}^2 F_{bc} + \dots~
\eea
in curved spaceo
To the best of our knowledge, such a result is not available in the literature. 


\section{Conclusion} \label{section6}

Building on Schwinger's work \cite{Schwinger:1951nm}, in this paper we have computed the one-loop 
Euler-Heisenberg actions for both spinor and scalar quantum electrodynamics in $d=2n>4$  dimensions, and derived the closed form expressions for these actions. We have also given the weak-field quantum corrections to the classical action in 6, 8 and 10 dimensions in terms of electromagnetic invariants.

The one-loop results for spinor and scalar QED in four dimensions \cite{Heisenberg:1936nmg, Weisskopf:1936hya, Schwinger:1951nm} were extended to two loops by Ritus in 1975 \cite{Ritus, Ritus:1977iu}  (see \cite{Dittrich:1985yb} for a review). Further two-loop analyses have been given by several groups using different techniques, see, e.g., \cite{BS, RSS,FRSS,KS}. It would be interesting to extend these two-loop results to six dimensions. 

Another interesting problem is to compute gravitational corrections to the Euler-Heisenberg Lagrangian in higher dimensions. 
In the $d=4$ case, this problem was studied some time ago in \cite{Bastianelli:2008cu}, see also a recent work \cite{Ahmadiniaz:2026yrw} for further developments and a complete list of references.
The $R F^2$ operators were computed in any dimension including $d=6$ via the heat kernel coefficients in \cite{Bittar:2024xuc}.

In six dimensions, the Lagrangian for conformal electrodynamics in curved space is a linear combination of the two composite primary dimension-6 fields \eqref{primary2} and \eqref{primary3}. It would be interesting to derive composite primary fields which determine the Lagrangians for conformal electrodynamics in eight and ten dimensions. We should emphasise that 
conformal Yang-Mills theories in $d=2n>6$ dimensions were constructed earlier  in Minkowski \cite{Metsaev:2023qif}
and anti-de Sitter \cite{Metsaev:2023qif} space, but Metsaev's construction was based on the use of auxiliary fields.

The impact of the work by Heisenberg and Euler \cite{Heisenberg:1936nmg} is comparable with that by Fradkin and Tseytlin \cite{Fradkin:1985qd} who derived the Born-Infeld action as  the effective action in the open bosonic string theory (see 
\cite{Metsaev:1987qp} for the supersymmetric extension and 
\cite{Tseytlin:1999dj} for the pedagogical  review).
Further aspects of (super)strings in a background electromagnetic field are described in 
\cite{Bachas:1992bh,Bachas:1996bp,Ambjorn:2000yr,Magnea:2004ai,Condeescu:2017nmo}.
The results of these publications are not directly related to those in our work, at least beyond ten dimensions.
\\

\noindent
{\bf Acknowledgements:}\\ 
We are grateful to Ian McArthur for useful comments and suggestions. 
This work is supported in part by the Australian Research Council, project DP230101629.


\appendix

\section{Spinor formalism in $d=2n$ dimensions}\label{appendixA}

This appendix consists of two parts. First of all, we briefly review the conceptual aspects of the spinor formalism in $d=2n$ dimensions. Secondly, we concentrate on the specific features of the $d=6$ case.  

\subsection{Gamma matrices in $d=2n$ dimensions}

In this subsection we follow
\cite{Gliozzi:1976qd, Scherk:1978fh}.
 Let $\g_a  $ be the gamma matrices in $d=2n$ dimensions, 
\begin{subequations} \label{gamma}
\bea
\{ \g_a , \g_b \} = -2 \eta_{ab} {\mathbbm 1}_{2^n}~, \qquad a,b= 0,1, \dots, d-1~,
\label{gamma.a}
\eea
with $\eta_{ab}$ the mostly plus Minkowski metric. We assume $\g_a$ to obey the standard Hermiticity condition
\bea
(\g_a)^\dagger = \g_0 \g_a \g_0 = -\eta^{ab} \g_b = - \g^a~, \qquad \g_a = (\g_0, \g_i)~.
\label{gamma.b}
\eea  
\end{subequations}
Let $\g_{d+1}$ denote 
the $d$-dimensional counterpart of the matrix $\g_5$ in four dimensions,
\begin{subequations}\label{gamma(d+1)}
\bea
\g_{d+1} = \eta \g^0 \g^1 \dots \g^{d-1}  ~, \qquad \eta^2 = (-1)^{(d-2)/2}~.
\eea
Its properties are
\bea
\g_{d+1} = (\g_{d+1})^\dagger~, \qquad (\g_{d+1})^2 ={\mathbbm 1}_{2^{n}} ~, \qquad {\rm tr} \,\g_{d+1} =0~,
\qquad \{ \g_{d+1} , \g_a\} =0.
\eea
\end{subequations}
We introduce the family of traceless matrices  
\bea
\g_{a(k)} := \g_{[a_1} \g_{a_2}\dots \g_{a_k]} ~, \qquad \g_{d+1} \g_{a(k)} = (- 1)^k \g_{a(k)} ~, \qquad
{\rm tr} \,\g_{a(k)} =0~,
\quad 1 \leq k \leq 2n~.
\eea
It holds that 
\bea
\g^b \g_{a(k)} \g_b = - (-1)^k (d-2k) \g_{a(k)}~. \label{UniversalRelation}
\eea

Of special importance in the spinor formalism are the  matrices $B$ and $C$ which solve of the equations 
\bea
(\g_a)^* &=& - B \g_a B^{-1}~, \label{B-matrix}\\
(\g_a)^{\rm T} &=& - C\g_a C^{-1} ~,
\label{C-matrix}
\eea
where $(\g_a)^* $ denotes the complex conjugate of $\g_a$.
Here $C$ is called the charge conjugation matrix. 
The matrices $B$ and $C$ can always be chosen to be unitary and they prove to 
have the symmetry properties:
\begin{subequations}
\bea
B^\dagger B &=& {\mathbbm 1}_{2^n} ~, \qquad B^{\rm T} = \e(d) B~, \label{B-matrix-symmetry}\\ 
C^\dagger C &=& {\mathbbm 1}_{2^n} ~, \qquad C^{\rm T} = -\e(d)C~, \label{C-matrix-symmetry}
\eea
\end{subequations}
where 
\begin{subequations} \label{epsilon(d)}
\bea
\e(d) = -\sqrt{2} \cos \left( \frac{\pi}{4} (d+1)\right)~, \qquad \e(d+8) = \e(d)~
\eea
or, explicitly,
\bea
\e(2)=\e(4)=1~,\qquad \e(6)= \e(8)=-1~,
\eea
\end{subequations} 
see \cite{Gliozzi:1976qd,Scherk:1978fh} for the derivation.
One can choose $C= B \g^0$. 

One can choose the Weyl representation of the gamma matrices  in which
\bea
\g_{d+1} = \left(\begin{array}{cc}{\mathbbm 1}_{2^{n-1}} ~&0 \\0~& -{\mathbbm 1}_{2^{n-1}}\end{array}\right)~,
\label{Weyl}
\eea
and therefore the matrices $\g_a$ are block off-diagonal, 
\bea
\g_a = \left(\begin{array}{cc}0~& (\s_a)_{\a \dot \b} \\
(\tilde{\s}_a)^{\dot \a \b} ~&0\end{array}\right)~.
\label{Weyl-gamma}
\eea
Here the sigma matrices obey the anti-commutation relations 
\bea
 \s_a \tilde{\s}_b + \s_b \tilde{\s}_a = - 2 \eta_{ab} {\mathbbm 1}_{2^{n-1}} ~,\qquad
\tilde{\s}_a {\s}_b + \tilde{\s}_b {\s}_a = - 2 \eta_{ab} {\mathbbm 1}_{2^{n-1}} ~ .
 \eea
 In the Weyl representation the Lorentz generators are block diagonal, 
 \bea \label{Lorentz}
 M_{ab} := - \frac 14 [\g_a , \g_b] = - \hf \g_{ab} 
 = -\hf \left(\begin{array}{cc}
 (\s_{ab})_\a{}^\b  ~& 0 \\
0 ~ &(\tilde{\s}_{ab})^{\dot \a }{}_{\dot \b} \end{array}\right)
=\big( (M_{ab})_{\hat \a}{}^{\hat \b}\big)~.
 \eea

 A Dirac spinor $\J$ has the form 
 \bea
 \J = \left(\begin{array}{c}\psi_{\a } \\
\bar \c^{\dot \a}
\end{array}\right) \equiv \big( \J_{\hat \a} \big)
\eea
and is characterised by the Lorentz transformation law
\bea
\J' = \exp \left(\hf \omega^{ab} M_{ab} \right)\J~, \qquad \L= \exp \,\o \in \sSO_0(d-1,1) ~, \qquad 
\o= (\o^a{}_b)~.
\eea
 The Dirac conjugate spinor $\bar{\J} = \J^\dagger \g^0$ transforms in the dual representation, 
 \bea 
  \bar \J' =  \bar \J \exp \left( - \hf \omega^{ab} M_{ab} \right)~.
 \eea
 
 
 \subsection{Gamma matrices in six dimensions}

In the $d=6$ case,\footnote{Our six-dimensional spinor formalism builds on  
\cite{Cederwall:2025ywy,Linch:2012zh, Butter:2016qkx}.}  
the charge conjugation matrix is symmetric, $C^{\rm T} = C$, in accordance with the relations
\eqref{C-matrix-symmetry} and
\eqref{epsilon(d)}. 
The symmetry of $C$ implies that 
\bea
\big(\g_{a(k)} C^{-1}\big)^{\rm T} =  (-1)^{\hf k(k+1)} \g_{a(k)} C^{-1}~.
\label{6Dsymmetry}
\eea
This relation tells us that the matrices $\g_a C^{-1}$ , $\g_{ab} C^{-1}$ , $\g_a \g_7 C^{-1}$ and $\g_{7} C^{-1}$ are anti-symmetric, while the matrices $C^{-1}$, $\g_{abc }C^{-1}$ and $\g_{ab} \g_{7} C^{-1}$ are symmetric, where $\g_7$ is the $d=6$ version of $\g_{d+1}$, eq. \eqref{gamma(d+1)}. We define the latter as
\bea
\g_7 := \g^0 \g^1 \dots \g^5~.
\eea
In the Weyl representation, eq. \eqref{Weyl}, it holds that $\{\g_7, C\}=0$. This means that the change conjugation
matrix is block off-diagonal, 
\bea
C = \left(\begin{array}{cc}0~& \mathfrak{c}^\a{}_ {\dot \b} \\
\mathfrak{c}_{\dot \a}{}^\b ~&0\end{array}\right)
= (C^{{\hat \a} {\hat \b}})
~, \qquad
\mathfrak{c}_{\dot \a}{}^\b = \mathfrak{c}^\b{}_{\dot \a}~,
\eea
compare with the gamma matrices \eqref{Weyl-gamma}.
It follows from the anti-symmetry of $\g_{ab} C^{-1}$ that the matrices $C^{-1} =(C_{\hat \a \hat \b})$ 
and $C=(C^{\hat \a \hat \b})$ are Lorentz invariant, 
\bea
M_{ab}C^{-1} + C^{-1} (M_{ab})^{\rm T} =0~,
\eea
and so is $\mathfrak{c}^\a{}_ {\dot \b} $ and its inverse. 
 
The Lorentz-invariant tensor  $ \mathfrak{c}^\a{}_ {\dot \b} $ and its inverse can be used to convert all dotted indices into undotted ones following the rules 
 \begin{subequations}\label{dotted-to-undotted}
\bea
\bar \c^{\dot \a} &\to& \bar \c^{\a} := \mathfrak{c}^\a{}_ {\dot \b}  \bar \c^{\dot \b} ~, \\
\qquad (\s_a)_{\a \dot \b} &\to& (\s_a)_{\a  \b} := (\s_a)_{\a \dot \g} (\mathfrak{c}^{-1})^{\dot \g}{}_\b~, \\
(\tilde{\s}_a)^{\dot \a \b} &\to & (\tilde{\s}_a)^{\a \b}
:= \mathfrak{c}^\a{}_ {\dot \g} (\tilde{\s}_a)^{\dot \g \b}~.
\eea
 \end{subequations}
As a result, one ends up with the gamma matrices
 \bea
{\g}{}_a = \left(\begin{array}{cc}0~& (\s_a)_{\a \b} \\
(\tilde{\s}_a)^{ \a \b} ~&0\end{array}\right)~,
\eea
and  for their off-diagonal blocks eq. \eqref{6Dsymmetry} implies
\bea
(\s_a)_{\a  \b} = - (\s_a)_{\b \a} ~, \qquad (\tilde{\s}_a)^{ \a \b} = - (\tilde{\s}_a)^{\b \a} ~.
\eea
For the matrices $\g_{ab}$, which determine the Lorentz generators in the spinor representation, \eqref{Lorentz}, we get
\bea
 \g_{ab} 
 =  \left(\begin{array}{cc}
 (\s_{ab})_\a{}^\b  ~& 0 \\
0 ~ &(\tilde{\s}_{ab})^{ \a }{}_{ \b} \end{array}\right)~, \qquad 
(\tilde{\s}_{ab})^{ \a }{}_{ \b} = -  (\s_{ab})_\b{}^\a~. 
\eea
We also have the important relations
\begin{subequations}
\begin{align}
     \tr(\sigma_{ab}\sigma_{cd}\sigma_{ef})F^{ab}F^{cd}F^{ef} &= -4\varepsilon_{abcdef}F^{ab}F^{cd}F^{ef} \,,\label{eq:sigmatrace1}\\
    \tr(\sigma_{ab}\sigma_{cd}) &= -4(\eta_{ac}\eta_{bd}-\eta_{ad}\eta_{bc}).\label{eq:sigmatrace2}
\end{align}
\end{subequations}
It also follows that the off-diagonal blocks of the matrices
 \bea
{\g}{}_{a(3)} = \left(\begin{array}{cc}0~& (\s_{a(3)})_{\a  \b} \\
(\tilde{\s}_{a(3)})^{ \a \b} ~&0\end{array}\right)
\eea
are symmetric,
\bea
(\s_{a(3)})_{\a  \b} =  (\s_{a(3)})_{\b \a} ~, \qquad 
(\tilde{\s}_{a(3)})^{\a \b}=  (\tilde{\s}_{a(3)})^{\b \a} ~.
\eea
  
 Let  $\varepsilon^{a_1 \dots a_6}$ be the completely antisymmetric Levi-Civita tensor normalised by  
 $ \varepsilon^{012345}=-1$. It holds that 
 \bea
\g^{a_1\dots a_k} \propto \widetilde{\g}^{a_1 \dots a_k} \g_{7}~,
 \qquad
 \widetilde{\g}^{a_1 \dots a_k} :=
\frac{1}{(6-k)!} \ve^{a_1 \dots a_k b_1 \dots b_{6-k} }
\g_{b_1 \dots b_{6-k}}
~,
\eea
Due to the identity
\bea
\g^{a(3)} = \frac{1}{3!} \ve^{a(3) b(3)} \g_{b(3)} \g_7~,
\eea
the matrices  $ {\s}_{a(3)}$ and $\tilde{\s}_{a(3)}$ are (anti) self-dual,
\bea
\frac{1}{3!} \ve^{a(3) b(3)} \s_{b(3)} = - \s^{a(3)} ~, \qquad
\frac{1}{3!} \ve^{a(3) b(3)} \tilde{\s}_{n(3)} =  \tilde{\s}^{a(3)}~.
\eea

In the $d=6$ case, the identity \eqref{UniversalRelation} implies the following:
\begin{subequations}
\bea
\s^b \tilde{\s}_{a(3)} \s_b =0~,
& \qquad & \tilde{\s}^b \s_{a(3)} \tilde{\s}_b =0 ~, \label{A.29a} \\
\s^b \tilde{\s}_{a} \s_b =4\s_a~,  &\qquad &
 \tilde{\s}^b \s_{a} \tilde{\s}_b = 4 \tilde{\s}_a~. \label{A.29b}
\eea
\end{subequations}
Introducing Lorentz-invariant tensors
\begin{subequations}
\bea
I_{\a\b, \g\d} &:=& (\s^b)_{\a\b} (\s_b)_{\g\d} = I_{[\a\b], [\g\d]} = I_{\g\d, \a\b}  ~, \\
\tilde{I}^{\a\b, \g\d} &:=& (\tilde{\s}^b)^{\a\b} (\tilde{\s}_b)^{\g\d} = \tilde{I}^{[\a\b], [\g\d]} = \tilde{I}^{\g\d, \a\b}  ~,
\eea
\end{subequations}
they allow us to raise and lower the spinor indices of $\s_a$ and $\tilde{\s}_a$,
\bea
\frac14 \tilde{I}^{\a \g, \d \b} (\s_a)_{\g\d} = (\tilde{\s}_a)^{\a\b}~, \qquad
\frac 14 {I}_{\a \g, \d \b} (\tilde{\s}_a)^{\g\d} = (\s_a)^{\a\b}~,
\label{sigma-tilde-sigma}
\eea
in accordance with \eqref{A.29b}. On the other hand, eq. \eqref{A.29a} tells us that the invariant tensors
$I_{ab, cd} $ and $\tilde{I}^{ab, cd} $ are completely antisymmetric,
\bea
I_{\a\b, \g\d} = I_{[\a\b, \g\d]}~, \qquad \tilde{I}^{\a\b, \g\d} = \tilde{I}^{[\a\b, \g\d]}~.
\eea

Let $\ve^{\a_1 \a_2 \a_3 \a_4} $ and $\ve_{\b_1 \b_2 \b_3 \b_4}$ be the spinor Levi-Civita tensor and its inverse, respectively,
 \begin{align}
 \ve^{\a_1 \a_2 \a_3 \a_4}  \ve_{\b_1 \b_2 \b_3 \b_4 }
 = 4! \d^{\a_1}{}_{ [\b_1} \d^{\a_2}{}_{\b_2} \d^{\a_3}{}\d_{\b_3} \d^{\a_4}{}_{\b_4]} ~.
 \end{align}
These tensors are used to raise and lower the antisymmetric rank-2 spinors associated with a six-vector $V^a$
\begin{subequations}
\bea
V^a ~\to ~ \tilde{V}{}^{\a_1 \a_2} := V^a (\tilde{\s}_a)^{\a_1 \a_2} ~, & \qquad &
V^a ~\to ~ {V}_{\a_1 \a_2} := V^a ({\s}_a)_{\a_1 \a_2} ~, \\
\tilde{V}{}^{\a_1 \a_2} = \hf \ve^{\a_1\a_2 \g_1 \g_2} V_{\g_1 \g_2} ~, &  \qquad &
 {V}{}_{\a_1 \a_2} = \hf \ve_{\a_1\a_2 \g_1 \g_2} \tilde{V}^{\g_1 \g_2} ~. 
\label{A18.b}
\eea
\end{subequations}
Comparing \eqref{A18.b} with \eqref{sigma-tilde-sigma} gives
\bea
\hf I_{\a_1\a_2, \a_3 \a_4} = \ve_{\a(4)}~, \qquad
\hf \tilde{I}^{\a_1\a_2, \a_3 \a_4} = \ve^{\a(4)}~.
\eea

The following completeness relations hold
\begin{subequations}
\bea
\d_\a{}^{[\g}\, \d_\b{}^{\d]} &=& \frac 14 (\tilde{\s}^a)^{\g\d} (\s_a)_{\a\b}~,  \\
\d_\a{}^{(\g}\, \d_\b{}^{\d)} &=& \frac{1}{3! \cdot 8} (\tilde{\s}^{a (3)})^{\g\d} (\s_{a(3)})_{\a\b}~.
\eea
\end{subequations}
Of special importance for us is the completeness relation
\bea
    \frac{1}{4}{(\tilde{\sigma}^{ab})^{\alpha}}_{\beta} {(\sigma_{ab})_\gamma}^\delta = -\frac{1}{2} {\delta^\alpha}_\beta {\delta^\gamma}_\delta + 2{\delta^\delta}_\beta {\delta^\alpha}_\delta\,. \label{eq:completeness}
\eea
It allows us to describe an antisymmetric second-rank tensors, $F_{ab} = -F_{ba}$, in terms of 
a traceless spinor matrix $\mathbb{F}= (F_\a{}^\b)$, $\tr \,\mathbb{F}=0$, by the rule:
\bea
    F_{ab} = \frac{1}{2}{(\sigma_{ab})_{\alpha}}^\beta {F_\beta}^\alpha ~ \iff ~{F_\alpha}^\beta = -\frac{1}{4}{(\sigma^{ab})_\alpha}^\beta F_{ab} \,.\label{eq:correspondence2}
\eea


\section{Electromagnetic invariants in six dimensions}\label{appendixB}

We now briefly discuss the invariant structure of electromagnetism in six-dimensions. The dual of $F_{ab}$ is a 4-form,
\bea
    \widetilde{F}^{abcd}=\frac{1}{2}\varepsilon^{abcdef}F_{ef}.
\eea
Both $F_{ab}$ and $F_{abcd}$ map to the same traceless matrix under the isomorphism described in \eqref{eq:correspondence2}. By the Cayley-Hamilton theorem, the spinor $\mathbb{F}$ satisfies its own characteristic equation,
\bea
    \mathbb{F}^4+c_3\mathbb{F}^3+c_2\mathbb{F}^2+c_1\mathbb{F}+c_0=0\,.\label{eq:spinorcharacteristic}
\eea
Higher powers than $\mathbb{F}^4$ can be written in terms of lower powers of $\mathbb{F}$. Since $\mathbb{F}$ is traceless we find that the independent invariants are quadratic, cubic and quartic in the field strength. A natural choice is,
\begin{subequations}\label{eq:invariants}
\begin{align}
    \mathcal{F}&=-\frac{1}{4}F_{ab}F^{ab}=\frac{1}{4}\tr\bm{F}^2\,,  \\
    \mathcal{G} &=\frac{1}{48}\varepsilon^{abcdef}F_{ab}F_{cd}F_{ef}=\frac{1}{24}\tilde{F}^{abcd}F_{ab}F_{cd}\,,\\
    \mathcal{H} &=\frac{1}{4}F_{ab}F^{bc}F_{cd}F^{da} = \frac{1}{4}\tr \bm{F}^4\,. 
\end{align}
\end{subequations}
We must also fix the coefficients $c_i$ in \eqref{eq:spinorcharacteristic} to find the eigenvalues $\omega_i$ of $\mathbb{F}$. We contract $\mathbb{F}^2$ with the completeness relation \eqref{eq:completeness} to obtain
\bea
    -\frac{1}{16}{(\tilde{\sigma}^{ab})^{\alpha}}_{\beta}\varepsilon_{abcdef}F^{cd}F^{ef} = -\frac{1}{2}\tr (\mathbb{F}^2){\delta^\alpha}_\beta + 2{(\mathbb{F}^2)_\beta}^\alpha .\label{eq:usefullater}
\eea
We then contract the same completeness relation with $\mathbb{F}^4$, and we use identities \eqref{eq:sigmatrace1}, \eqref{eq:sigmatrace2} along with \eqref{eq:usefullater} to simplify. We obtain the result
\bea\label{eq:spinorcharacteristic2}
    \mathbb{F}^4-\mathcal{F}\mathbb{F}^2-\mathcal{G}\mathbb{F}+\frac{1}{4}(\mathcal{H}-\mathcal{F}^2)\mathbbm{1}=0.
\eea
In order to compute,
\bea
\sqrt{\det\left(\frac{e\bm{F}s}{\sin{e\bm{F}s}}\right)}~,
\eea
we also need to know the eigenvalues of the field strength tensor. These can be calculated using the formalism discussed in Section \ref{section4}. In six dimensions, the characteristic equation is of the form,
\bea
\l^6 + e_2\l^4+e_4\l^2 - \mathcal{G}^2 = 0~,
\eea
where $e_n$ is the elementary symmetric polynomial in $\l$ of order $n$. Newton's identities give,
\begin{subequations}
\bea
e_2 &=& -\frac{1}{2}\tr(\bm{F}^2) = -2\mathcal{F}~,\\
e_4 &=& \frac{1}{8}\tr(\bm{F}^2)^2-\frac{1}{4}\tr(\bm{F}^4) = 2\mathcal{F}^2-\mathcal{H}~.
\eea
\end{subequations}
We find the characteristic equation,
\bea\label{eq:tensorcharacteristic}
    \lambda^6-2\mathcal{F}\lambda^4+(2\mathcal{F}^2-\mathcal{H})\lambda^2 -\mathcal{G}^2=0~.
\eea
From this, Vi\`ete's formulas imply
\begin{subequations}\label{eq:vieta}
\begin{align}
    \lambda_1^2+\lambda_2^2+\lambda_3^2 &= 2\mathcal{F} ~,\\
    \lambda_1^2 \lambda_2^2+\lambda_1^2\lambda_3^2+\lambda_2^2\lambda_3^2&=2\mathcal{F}^2-\mathcal{H}~,\\
    \lambda_1^2\lambda_2^2\lambda_3^2 &= \mathcal{G}^2 ~.
\end{align}
\end{subequations}
For a more in-depth discussion on the structure of invariants in diverse dimensions, see \cite{Cederwall:2025ywy}.


\section{DeWitt's  coefficients} \label{appendixC}
In this section, we briefly review the Schwinger–DeWitt technique \cite{DeWitt:1964mxt,DeWitt:2003pm} and use it to compute the $\tr[a_3]$ heat kernel coefficient for spinor QED in six dimensions, as well as the $[a_3]$ and $[a_4]$ coefficients for scalar QED. These coefficients are local invariants constructed from the background fields and their derivatives, and they determine the one-loop effective action. DeWitt's method provides a systematic procedure for computing the heat kernel coefficients associated with minimal (Laplace-type) operators in flat or curved spacetime. Here, we restrict our attention to massive fields in flat spacetime, for which the relevant operators take the form
\bea
\D = -\square+m^2\mathbbm{1}+\hat{P}~,\quad \square =\nabla^a\nabla_a~.
\eea
Such operators are present in equations \eqref{eq:effectiveactiondef} and \eqref{scalareffectiveaction}.
$\hat{P}$ is an endomorphism and may be a matrix or a scalar depending on the field theory in question. The heat kernel admits the well-known ansatz,
\bea
    K(x,x';s) = \frac{\ri}{(4\pi \ri s)^{d/2}}\re^{\frac{\ri}{2s}\sigma(x,x') - \ri m^2 s}\Lambda(x,x';s)~,
    \quad  \sigma(x,x') 
    = \frac{1}{2}(x-x')^2~.
    \label{eq:heatansatz}
\eea
By inserting \eqref{eq:heatansatz} into \eqref{eq:definingrelation}, one finds the equation
\bea
-\frac{\pa{\Lambda}}{\pa s}+\ri\square\Lambda =\frac{\ri}{s}\partial^a\sigma\nabla_a\Lambda+\ri \hat{P}\L~.\label{1}
\eea
We are interested in the small proper-time behaviour of the effective action. We can expand the function $\L(x,x';s)$ as a power series in the proper time,
\bea
    \Lambda(x,x';s) = \sum_k a_k(x,x')(\ri s)^k~ .\label{eq:powerseries}
\eea
We insert the power series expansion \eqref{eq:powerseries} into \eqref{1} and demand that terms with different powers of the proper-time $s$ vanish independently. This yields two relations,
\begin{subequations}
\begin{align}
\partial^a\sigma \nabla_a a_0&=0
    \label{3}~,\\
    (k+1)a_{k+1}+\partial^a\sigma \nabla_a a_{k+1}&=\square a_k-\hat{P}a_k~. \label{recursion}
\end{align}
\end{subequations}
Equation \eqref{3} implies that $a_0(x,x')$ is the parallel displacement operator,
\bea
    a_0(x,x') = \mathcal{I}(x,x')~, \quad [\mathcal{I}]:=\mathcal{I}(x,x) = \mathbbm{1}~,
\eea
while equation \eqref{recursion} is a recursion relation for $a_k$.  We use square brackets to denote coincidence limits. Physical information about the one-loop effective action is carried by the heat kernel at the coincidence limit, and so the objects that we are interested in are $[a_k]=a_k(x,x)$. DeWitt's original method does not allow us to obtain closed-form expressions for the coefficients $a_k(x,x')$ beyond coincidence. However, they can be reconstructed from the coincidence limits $[a_k]$ using a covariant Taylor series expansion. The off-diagonal components $a_k(x,x')$ are not of interest to us, but they are of relevance in the computation of higher-loop contributions or for non-minimal operators (for example, see \cite{Barvinskii:2024iqz}).

If we differentiate equation \eqref{recursion} and take coincidence limits, we can express terms of the form $[\nabla^{a_1}\nabla^{a_2}\dots\nabla^{a_i}a_{k+1}]$ in terms of $[\nabla^{a_1}\nabla^{a_2}\dots\nabla^{a_j}a_k]$, where $i\leq j+2$. By sequentially calculating these quantities, we can obtain the exact local expressions for $[a_k]$ in terms of contractions of derivatives with the background fields. 

\subsection{$a_3$ coefficient in six-dimensional spinor QED}
We begin with six-dimensional spinor electrodynamics. The endomorphism $\hat{P}$ is
\bea
\hat{P} = -\frac{\ri e}{2}\g^{ab}F_{ab}~,
\eea
and so the recursion relation takes the form
\bea
    (k+1)a_{k+1}+\partial^a\sigma \nabla_a a_{k+1}=\square a_k+\frac{\ri e}{2}\g^{ab}F_{ab}a_k~. 
\eea
The results for $a_1$ and $a_2$ are well known, but we briefly show them here to illustrate the method. To compute $a_1$, we set $k=0$ in \eqref{recursion} and take a coincidence limit, which gives
\bea
    [a_1] =\ri [\square\mathcal{I}] + \frac{\ri e}{2}\gamma^{ab}F_{ab}~.
\eea
In order to compute $[\square\mathcal{I}]$, we repeatedly differentiate \eqref{3}. Taking a coincidence limit yields $[\square\mathcal{I}]=0$. We obtain
\bea
    [a_1]=\frac{\ri e}{2}\gamma^{ab}F_{ab}~.
\eea
Since the relevant differential operator for spinor QED is matrix valued, the functional trace present in \eqref{eq:effectiveactiondef} is also a trace over spinor indices. The relevant contributions to the effective action are thus $\tr[a_k]$. We find that $\tr[a_1]=0$, as $\gamma^{ab}$ is traceless. For $[a_2]$, we find
\bea
[a_2] = \frac{1}{2}[\square a_1]-\frac{1}{4}e^2\g^{ab}\g^{cd}F_{ab}F_{cd}~.
\eea
We can compute the trace over $\g^{ab}\g^{cd}$ in six dimensions using the formalism presented in Appendix \ref{appendixA}. Computing $[\square a_1]$ from the recursion relation and evaluating the trace yields
\bea
\tr[a_2]=-\frac{16}{3}e^2\mathcal{F}~.
\eea
The procedure for $[a_3]$ is much the same, but more tedious. In total, we find
\begin{subequations}
\begin{align}
    \tr[a_0] &= 8 ~,\\
    \tr[a_1] &= 0 ~,\\
    \tr[a_2] &=  -\frac{16}{3}e^2\mathcal{F}~,\\
    \tr[a_3] &= \frac{2}{5}e^2F^{ab}\square F_{ab}+\frac{7}{45}e^2 \partial^a F_{ba}\partial_c F^{bc}-\frac{2}{45}e^2\partial^a F^{bc}\partial_a F_{bc} ~.\label{eq:traces}
\end{align}
\end{subequations}
In the computation of $\tr[a_3]$, the Bianchi identity is also used to  permute derivatives of the field strength. 

The coefficient $[a_3]$ controls the logarithmic divergence of the
effective action in six dimensions and leads to a higher-derivative
counterterm proportional to $F^{ab}\square F_{ab}$. When integrated by parts, $\tr[a_3]$ becomes
\bea
    \tr[a_3]=\frac{11e^2}{30}F^{ab}\square F_{ab} + \text{total derivative}~,
\eea
which is of the form \eqref{6d-action}. 

\subsection{$a_3$ and $a_4$ coefficients in scalar QED}
For scalar electrodynamics, the endomorphism $\hat{P}=0$, and so the recursion relation \eqref{recursion} becomes
\bea
(k+1)a_{k+1}+\partial^a\sigma \nabla_a a_{k+1}=\square a_k~,\label{scalarrecursion}
\eea
where $a_k$ are scalars. We perform an essentially identical calculation, made slightly simpler by the fact that there is no trace over spinor indices. The first three Seeley-DeWitt coefficients in this case are,
\begin{subequations}
\begin{align}
    [a_0]&=1~,\\
    [a_1]&=0~,\\
    [a_2]&=\frac{1}{3}e^2\cal{F}~,\\
    [a_3]&=-\frac{1}{30}e^2F^{ab}\square F_{ab} - \frac{1}{45}e^2\partial^a F^{bc}\partial_a F_{bc}-\frac{1}{180}e^2 \partial^a F_{ba}\partial_c F^{bc}\,. \label{a3-scalar}
\end{align}
\end{subequations}
Upon integrating by parts,
\bea
    [a_3]=-\frac{e^2}{120}F^{ab}\square F_{ab}~.
\eea  
Our result for $[a_3]$ agrees with the earlier calculation  given in \cite{Bastianelli:2000hi}.

The computation of the $a_4$ coefficient for scalar QED is quite a bit more involved. This object is a conformally invariant Lagrangian for $8$-dimensional electrodynamics. Using the recursion relation \eqref{scalarrecursion} we find
\bea
[a_4] = \frac{1}{120}[\square^3 a_1]+\frac{1}{30}\ri eF^{ab}[\nabla_b\nabla_a a_2]+\frac{1}{60}\ri e\partial_bF^{ab}[\nabla_a a_2]~,
\eea
where
\bea
[\square^3 a_1] = \frac{1}{3}[\square^4 \mathcal{I}]-\frac{4}{3}\left([\square\nabla^a\square\nabla a_1]-\ri eF^{ad}[\nabla_d\square \nabla_a a_1]-\frac{1}{2}\ri e\partial_dF^{ad}[\square \nabla_a a_1]\right)~.
\eea
For brevity, it is convenient to specify the following basis of dimension +8 operators,
\begin{align*}
&I_1 =  e^4(\tr{\bm{F}^2})^2, \quad I_2 = e^2F_{ba}\square^2 F^{ba}~, \quad I_3 = e^2\square F_{ba}\square F^{ba}~,\quad I_4=e^2\partial^c F^{ba}\square\partial_c F_{ba}~,\\ \quad &I_5 = e^2\partial^d\partial^cF^{ba}\partial_d\partial_cF_{ba}~,\quad I_6 = e^2\partial_a F^{ba}\square\partial^d F_{bd}~, \quad I_7 = e^2\partial_c\partial^bF_{bd}\partial^c\partial_aF^{ad}~, \quad I_8 = e^4\tr(\bm{F}^4)~.
\end{align*}
In terms of these, we find,
\bea
[a_4] = \frac{1}{288}I_1-\frac{1}{280}I_2 - \frac{1}{336}I_3 - \frac{1}{105}I_4 - \frac{1}{280}I_5 - \frac{1}{900}I_6-\frac{1}{1260}I_7+\frac{1}{360}I_8~.
\eea
In the constant-field regime, all terms with derivatives vanish and this expression reduces to
\begin{align*}
[a_4] &= \frac{1}{288}I_1+\frac{1}{360}I_8\\
&= \frac{e^4}{90}(5\mathcal{F}^2+\mathcal{H})~,
\end{align*}
which agrees with the quartic weak-field correction \eqref{quarticwf} prior to computing the proper-time integral. Upon integrating by parts, we find,
\bea
[a_4] = \frac{e^4}{288}(\tr{\bm{F}^2})^2+\frac{e^4}{360}\tr(\bm{F}^4) -\frac{11e^2}{25200}F_{ab}\square^2F^{ab} +\text{total derivative}~.
\eea
This Lagrangian describes a conformally invariant Lagrangian in $8$ dimensions and 
 involves  the same structures as in eqs. \eqref{6d-action} and \eqref{8d-action-b}. 

The DeWitt's coefficients $[a_3]$ and $[a_4]$ were computed in 1988 in \cite{Amsterdamski:1988bc} for the case of spinor QED in four dimensions.


\begin{footnotesize}

\end{footnotesize}

\end{document}